\documentclass[12pt,preprint]{aastex}
\begin {document}
\def\3he{$^3$He}
\def\4he{$^4$He}
\def\6li{$^6$Li}
\def\7li{$^7$Li}
\def\he3{$^3$He}
\def\eg{{\it e.g.}}
\def\ie{{\it i.e.}}
\def\etal{{\it et al.\ }}
\title{Stellar Mixing and the Primordial Lithium Abundance}
\author{M.H. Pinsonneault \altaffilmark{1},
G. Steigman \altaffilmark{1,2},
T.P. Walker \altaffilmark{1,2},
\& V.K. Narayanan \altaffilmark{3}}
\authoraddr {The Ohio State University, Columbus, OH 43210}
\altaffiltext {1} {Department of Astronomy, The Ohio State Univerity}
\altaffiltext {2} {Department of Physics, The Ohio State University}
\altaffiltext {3} {Princeton University, Dept. Astrophysical Sciences
and Observatory}
\date{January 11 2001}
%

\begin{abstract}
We compare the properties of recent samples of the lithium abundances in 
halo stars to one another and to the predictions of theoretical 
models including rotational mixing, and we examine the data for 
trends with metal abundance.  We apply two statistical tests to 
the data: a KS test sensitive to the behavior around the sample 
median, and Monte Carlo tests of the probability to draw the 
observed number of outliers from the theoretical distributions.  
We find from a KS test that in the absence of any correction for
chemical evolution, the \cite{RNB}  (hereafter 
RNB) sample is fully consistent with mild rotational mixing induced 
depletion and, therefore, with an initial lithium abundance higher 
than the observed value.  Tests for outliers depend sensitively on 
the threshold for defining their presence, but we find a 10$--$45\% 
probability that the RNB sample is drawn from the rotationally 
mixed models with a 0.2 dex median depletion with lower probabilities 
corresponding to higher depletion factors.  When chemical evolution 
trends (Li/H versus Fe/H) are included in our analysis we find that 
the dispersion in the RNB sample is not explained by chemical 
evolution; the inferred bounds on lithium depletion from rotational 
mixing are similar to those derived from models without chemical 
evolution.  Finally, we explore the differences between the RNB 
sample and other halo star data sets.  We find that differences 
in the equivalent width measurements are primarily responsible 
for different observational conclusions concerning the lithium 
dispersion in halo stars.  Implications for cosmology are discussed.  
We find that the standard Big Bang Nucleosynthesis predicted 
lithium abundance which corresponds to the deuterium abundance 
inferred from observations of high-redshift, low-metallicity 
QSO absorbers requires halo star lithium depletion in an amount consistent 
with that from our models of rotational mixing, but inconsistent with
no depletion.
\end{abstract}

\keywords{stars: abundances; cosmology: cosmological parameters;
stars: rotation}

\section{Introduction}
The primordial abundance of the light element lithium provides a
crucial test of Big Bang nucleosynthesis (BBN); it is also an important
diagnostic of standard and nonstandard stellar evolution theory.
The detection of \7li in halo stars by \cite{Spites} opened 
up the prospect of the direct detection of the primordial lithium
abundance.  There have been a number of subsequent observational
efforts which have produced a detailed picture of the distribution 
of halo star lithium abundances (\cite{Sp93}; \cite{T94}
(hereafter T94); \cite{BM97}; RNB; see also \cite{Ryan96}).
Primordial \7li, as one of the four light nuclides produced
in measurable abundance in standard BBN (the others being D, 
\3he, and \4he $--$ see \cite{OSW} for a review), provides a crucial 
consistency check in that all four nuclides are determined by the 
one free parameter of standard BBN $--$ the baryon-to-photon ratio, 
$\eta$.  Currently, the primordial deuterium abundance provides 
the best estimate of $\eta$.  However, the BBN-predicted primordial 
lithium abundance which is consistent with the observationally 
inferred primordial deuterium abundance (and thus with our best 
estimate of $\eta$) is actually {\it much} larger than the lithium
abundance observed in halo stars.  We show that theoretical models
which include rotational mixing (and are required by the observed
dispersion of halo lithium abundances) predict a primordial lithium
abundance which is consistent, in the context of standard BBN, with
the observed primordial deuterium abundance.

\subsection{Stellar Models Compared with Earlier Data Sets}

The interpretation of the halo star data requires knowledge of 
the stellar evolution effects which have influenced the surface 
abundance during the lifetime of the stars.  In ``classical'' 
(\ie, nonrotating) stellar models lithium is destroyed on the main 
sequence only in the presence of a deep surface convection zone; 
some pre-main sequence depletion will occur for a wider range of 
masses.  Such models predict only small amounts of lithium depletion 
for the hottest subdwarfs (effective temperature greater than about 
5800 K) and for their Population I analogs (\eg, \cite{DDK90}).  
In the Population I case classical models make detailed predictions 
about lithium depletion which can be tested using 
data from open clusters with a range of ages.  The open cluster 
data is in strong contradiction with the predictions of classical 
models.  In particular, there is observational evidence for a 
dispersion in lithium abundance at fixed mass, composition, and 
age, and also for lithium depletion on the main sequence in stars 
with surface convection zones too shallow to burn lithium in the 
classical models (e.g. \cite{Bal95}; \cite{MP97}; \cite{JFS99}).  
The rate of main sequence depletion is observed to decrease with 
age, and there are also strong mass-dependent depletion effects, 
none of which are predicted by the classical stellar models.

A number of physical mechanisms neglected in classical stellar 
models have been suggested as possible causes for the discrepancies.  
Rotational mixing is one attractive explanation since a range 
in initial rotation rates will produce a range in rotational 
mixing rates and the rate of rotational mixing would decrease 
with age as low mass stars lose angular momentum.  Unfortunately, 
models of Population II stars cannot be subjected to the same 
stringent level of tests that can be performed for open cluster 
stars.  The major unique signature of rotational mixing in the 
Population II context is therefore the presence of a {\it dispersion} 
in lithium abundance at fixed mass and composition.  As a result, 
much of the theoretical and observational work on the subject has 
therefore focused on the existence and magnitude of dispersion 
in the Population II lithium abundances.

In a previous paper (\cite{PWSN}, hereafter PWSN) we computed 
the distribution of \7li depletion factors expected from stellar
models including rotational mixing.  The distributon of depletion
factors was compared with the largest uniform data set available, 
that of T94.  We concluded that a combination of the observed
dispersion in abundances, the relative depletion of the isotopes 
\7li and \6li, and the existence of a small population of highly 
depleted stars all argued in favor of the stellar depletion of 
lithium and we placed bounds of 0.2 $--$ 0.4 dex on the \7li depletion
factor.  In this paper we compare our theoretical calculations 
with the newer halo lithium data set of RNB.  

The principal properties of lithium depletion in stellar models 
which include rotation can be summarized as follows.  Rotation 
can induce mixing in the radiative interiors of stars leading to 
surface lithium depletion during the main sequence phase of evolution.  
This depletion due to rotational mixing is in addition to surface 
lithium depletion during the pre-main sequence and (in the case 
of cool stars) main sequence evolution.  The degree of rotational 
mixing depends on the angular momentum content and its evolution 
so that a range of pre-main sequence rotation rates will produce a 
range of lithium depletion factors, in the sense that rapid rotators 
experience more mixing and lithium depletion than do slow rotators.  
There is compelling evidence from the Population I data for main 
sequence lithium depletion as well as for a dispersion in lithium 
abundance at fixed mass, composition, and age; rotational mixing 
naturally explains this pattern.  PWSN found that halo star models 
experience systematically less lithium depletion than do solar 
abundance models for the same sets of initial conditions.

The distribution of lithium depletion factors depends on the distribution 
of initial conditions, which can be inferred for young Population I 
clusters.  The distribution of pre-main sequence rotation rates 
needed to reproduce the rotation data in the Pleiades cluster 
produced a degree of dispersion which is correlated with the 
absolute amount of \7li depletion.  Since the majority of young
stars have similar rotation rates, the majority of stars will 
experience similar \7li depletions.  There is, however, a 
subpopulation of rapid rotators that are predicted to experience 
higher \7li depletion.  Comparison with the T94 data set led to 
a range of 0.2 $--$ 0.4 dex in the inferred stellar depletion (PWSN).  
When combined with an observed ``Spite plateau'' \7li abundance 
of 2.25 $\pm$ 0.10 (on the logarithmic scale where H = 12.0), this 
yielded a primordial \7li abundance in the range of 2.35 $--$ 2.75.
We emphasize that the PWSN models have the following overall
properties: in contrast to a simple gaussian distribution of 
abundances there is a distribution with a core whose dispersion 
is dominated by observational errors, along with a subpopulation 
(of order 1/5 of the sample) with moderately higher depletion 
factors, and a smaller population of (order 2-3\% of the sample) 
with large depletion factors.  These features will prove important 
in our comparison with newer halo star data of RNB.

\subsection{New Results from RNB}

The PWSN conclusions have recently been challenged by RNB using 
data from a high precision study of lithium abundances in a smaller, 
albeit still significant, sample of halo stars.  They obtained 
both a lower absolute observed abundance (2.11) and a significantly 
reduced error estimate and dispersion.  They attributed the residual 
dispersion to chemical evolution (e.g. the observed spread in Li/H 
in their view is caused by differences in post-BBN lithium production 
correlated with the range in Fe/H).  They argue 
that their data set {\it requires} stellar depletion be minimal.  In 
recent papers (\cite{Ryan00}, Suzuki, Yoshii, \& Beers 2000) 
the RNB results have been used to argue that the primordial lithium 
abundance is below 
their observed value in very metal-poor stars as a result of galactic 
production.  In this paper we compare our models with this new data 
set and we also compare the RNB data set with other studies.  We 
begin by comparing the data set of RNB with the theoretical 
distributions of PWSN in section 2.  We analyze the dispersion and 
chemical evolution trends in section 3 and compare the RNB and T94 
data sets in section 4.  Our conclusions concerning the primordial 
abundance of lithium and its consequences for cosmology are found 
in section 5.

\section{Comparison of the Models with the Data}

The RNB sample does have a dispersion in excess of their observational 
errors.  RNB attribute this excess dispersion to chemical evolution.  
We will begin by comparing the RNB data to theory without any chemical 
evolution detrending; we consider both the reality of any trend with 
metallicity and its impact on the inferred lithium abundance in section 
3.  We note here that none of the overall conclusions of the comparison 
between data and theory in this section are dramatically modified by 
the treatment of chemical evolution effects (see section 3.)  
Furthermore, an analysis of the models without metallicity 
detrending provides the least model-dependent constraint on 
the degree of rotational mixing.

Differences in stellar rotation rates will produce differences 
in the degree of rotational mixing, so excess dispersion can be 
a signature of stellar depletion.  However, the distribution of 
stellar rotation rates is needed in order to predict the distribution 
of stellar lithium depletion factors.  Stellar models with rotation 
must also account for angular momentum loss from a magnetic wind 
and internal angular momentum transport.  Finally, the degree of 
mixing for a given angular momentum distribution must be specified 
(see PWSN for a more detailed description.)

As discussed in PWSN, rotation data in young open cluster stars 
is our best current guide to the initial conditions that might be 
applicable to halo stars.  The majority of young stars are slow 
rotators with similar rotation rates; these stars will have almost 
uniform depletion and very little internal scatter.  About 15\% 
of young stars are rapid rotators, including a subpopulation (about 
3\%) of very rapid rotators.  This will produce a tail of overdepleted 
stars in the distribution.  There are unavoidable observational 
selection effects which may influence the inferred distribution 
of rotation velocities.  For example, very slow initial rotators 
would only have upper limits to their rotation velocity, so it 
is difficult to estimate how many stars should be underdepleted 
compared to the median.  There are occasional claims of pre-MS 
stars with very long periods, and this might explain the occasional 
halo star above the lithium plateau.  At the other end, the rapid 
rotator tail is subject to Poisson noise $--$ the fastest spinner in 
the Pleiades is at 140 km/s and the second fastest is at 90 km/s.  
So the very far tail of the underdepleted stars is difficult to 
pin down.  However, the behavior of the peak of the distribution 
is not sensitive to these details.  This provides justification 
for our including the one upper limit lithium abundance in the 
RNB sample and considering those outliers below, but not above, 
the median in our tests of the models.

We can empirically constrain the angular momentum loss and transport
properties by comparing different classes of theoretical models to
stellar observations as a function of mass and age.  Angular momentum
transport and mixing by hydrodynamic mechanisms are included in the
models.  We calibrate the mixing by requiring that a solar model
reproduce the solar lithium depletion at the age and rotation rate of
the Sun.  However, we have no direct information on the solar initial
conditions; because angular momentum loss scales as $\omega^3$, stars 
with a wide range of initial rotation rates end up with similar
rotation rates at old ages.  In PWSN, we considered three solar calibrations
(s0, s0.3, and s1) which correspond to three different overall
normalizations for the stellar lithium depletion.  The s0 case assumes
that the Sun was initially a rapid rotator, so the typical star will
experience much less depletion than the Sun; the s0.3 and s1 cases
correspond to assuming the Sun is more typical and the overall
expected stellar depletion is therefore larger.

The \cite{PDD92}
depletion factor of 10 from rotational mixing came from the assumption
that the Sun was a typical star; furthermore, these early models did
not include a saturation of angular momentum loss for rapidly rotating
stars.  The current generation of models is in significantly better
agreement with more recent measurements of stellar rotation rates,
which both permits us to infer the distribution of rotation rates and
to rule out depletion factors as large as the 1992 values.

There will be two principal differences between rotationally mixed and
standard models that can be directly tested with the halo star lithium
data.  The internal range in rotation among slow rotators will produce
an increase in the dispersion around the sample median relative to the
observational errors; and the rapid rotators will be overdepleted
relative to the median.  We therefore apply two statistical tests to
the data.  We compare the cumulative distribution of stars to
theoretical distributions anchored at the median abundance of the
sample using a KS test.  This test allows us to measure the
constraints on stellar depletion from the tightness of the bulk of the
halo lithium plateau stars.

We also applied both a simple analytical model and Monte Carlo
simulations to test the probability of drawing the observed number 
of outliers from the theoretical simulations.  For a given distance 
below the median there is a probability that any given star in the 
theoretical distribution will lie at or below that abundance ($P_{0}$).  
For a sample of size N, the probability that there will be a given
number I of stars below such a threshold is $S_{I}P_{0}^{I}(1-P_{0})^{N-I}$
where $S_{I}$ is the number of states capable of producing a given
number of outliers.  For I = 0,1, $S_{I}$ = 1,N respectively; it is
straightforward if tedious to compute the number of accessible states
for more outliers.  We used Monte Carlo simulations to check for the
cases with larger numbers of outliers.

We also compare to a Gaussian distribution of errors.  This permits us
to test for the possibility that the excess dispersion arises from a 
global underestimate of the observational errors and to quantify the
relative agreement of models with and without stellar depletion.

\subsection{Comparison with the Cumulative Distribution}

We convolved the theoretical distributions for the s0, s0.3, and s1
cases of PWSN described above with 
gaussian observational errors of 0.035 dex (see
Section 4 for our determination of the observational errors).  
In Figure 1 we compare the cumulative RNB distribution with 
these three models and a Gaussian distribution of errors.  The 
s0, s0.3, and s1 models have
median depletion factors of 0.18 dex, 0.32 dex, and 0.50 dex
respectively; on the RNB abundance scale these would correspond to
initial abundances of 2.29, 2.43, and 2.56 respectively.  Despite the 
very different depletion factors, all 
of the models have similar properties in the core of the distribution.  
Only the s1 case, with a high median stellar depletion of 0.50 dex, 
predicts a core broader than the observed distribution.
Gaussian errors alone cannot reproduce both 
the tightness of the core and the presence of outliers in the sample.

KS tests applied to this distribution indicate that there is, 
respectively, a 60\%, a 25\%, and a 5\% chance that the s0, s0.3, 
and s1 cases could be drawn from the same distribution as the data; 
by comparison the Gaussian has an 86\% chance of being drawn 
from the same distribution as the data.  We conclude from this test, 
which is primarily sensitive to the behavior of a sample around the 
median, that only high depletion factors are problematic $--$ although 
even the highest depletion case is only excluded at the 95\% level.
This confidence level is too low to absolutely rule out a model.
There is thus no contradiction between mild stellar depletion
from rotational mixing and the presence of a core of halo lithium
abundances with small internal scatter.  This data
suggests that the internal dispersion in the core is primarily caused
by observational error, and furthermore that an underestimate of the
observational errors is not responsible for the excess scatter in the data.
At the same time, it also provides no support for depletion factors at
or above the 0.5 dex level.

\subsection{Number of Outliers}

Inspection of Figure 1 reveals that the theoretical distributions 
predict more overdepleted stars than are present in the data set, 
but that there are more overdepleted stars in the sample than are
predicted by the observational errors (compare the solid and 
dotted curves).  Because of the small number of stars (23) in the 
sample, and the even smaller number of outliers expected in the 
sample (3 $--$ 7), we believe that only tentative claims can be made 
about the consistency (or lack thereof) of modest depletion factors 
with the data.  The basic issue is simply that the expected number 
of outliers in the rotational mixing models is small for a sample 
of 23 stars, which makes the conclusions subject to Poisson noise.

As an illustration, consider the different distributions in Figure 
1.  The tail of the observed distribution up to an abundance of 2.0 
corresponds to three overdepleted stars; it is clearly inconsistent 
with the expectation from observational errors, since abundances this 
low are formally three $\sigma$ or more below the median and therefore 
very unlikely in a sample of 23 stars.  The highest depletion case 
predicts more stars more than 0.1 dex below the median (7) than are 
observed (3).  However, the sample is so small that the specific 
statistical conclusions depend sensitively on where the threshold 
for defining an outlier is defined.  If the threshold is defined 
at 2.01 (just above two of the three overdepleted stars), then 
the expected fraction of outliers relative to the data is minimized 
and there is a 45\% chance of drawing the observed number of stars
relative to the s0 case and less than a 0.1\% probability of seeing 
as many as three outliers from observational errors alone.

However, there is a gap in the sample between abundances of 2.00 
and 2.06; if an outlier is defined as being at or below 2.05 the 
expected outlier fraction is increased and the observed outlier 
fraction is the same.  In this case there is a 10.6\% chance of 
drawing the data from the minimally depleted s0 distribution and 
an 11.4\% chance of seeing as many as three outliers from observational 
errors.  Similar fluctuations arise from excluding the one upper 
limit from the sample or clipping the tail of the theoretical 
distribution.  We therefore consider a range of probabilities 
from the most stringent (counting all stars more than 0.06 dex 
below the median as outliers) to the least stringent (counting 
all stars more than 0.10 dex below the median as outliers.)

\begin{deluxetable} {llll}
\tablecaption{Outlier Tests, No Evolution}
\tablehead{
\colhead{Case} & \colhead{$log (Li/Li_o)$} & \colhead{Probability} &
\colhead{Probability} \\
 & & \colhead{$<2.06$} & \colhead{$<2.01$} 
}
\startdata

Gaussian & 0.00 & 0.114 (1.0) & 0.00005 (0.02) \\
s0 & -0.18 & 0.45 (3.7) & 0.11 (5.5) \\
s0.3 & -0.32 & 0.15 (5.1) & 0.034 (7.4) \\
s1 & -0.50 & 0.05 (6.9) & 0.014 (8.3) \\
\enddata
\end{deluxetable}

The numbers in parenthesis after the 
listed fractional probabilities in the second and third columns of Table 1 are the expected number of outliers if
we set the threshold for defining one at less than 2.06 or less than
2.01 respectively.  The actual number of outliers is three below 2.01
or below 2.06, \eg,  there are no stars between 2.00 and 2.06.  The
probabilities for the Gaussian are for having three or more outliers;
the probabilities for the other three cases are for having three or
fewer outliers.  Because of
sparse sampling there is a range of possibilities for defining what an
outlier is.  The closer the cut is to the median, the larger the
number of expected outliers; this favors the no-depletion case
because there are more outliers than expected, but disfavors the stellar
depletion case because there are fewer outliers than expected.

If we exclude the one upper limit (G186-26), the minimum/maximum probabilities
for the s0 case drop to 5.5\% and 24.8\% respectively for two outliers out
of 22 stars.  \cite{Ryan01} have argued that the
rare ultra-lithium depleted stars are binary merger products, and that
they should therefore be excluded from samples of this type.  In
support of this they note that there is a large difference in
abundance between the ultra-depleted stars (of order 5\%) and others
and that the fraction of overdepleted stars is very high in
intermediate metal abundance stars which are hot enough to be
plausible blue straggler candidates.

We first note that \cite{Ryan01} does not establish a causal link between
high lithium depletion and binary merger products; in fact, the
authors argue that excess lithium depletion may be the sole indicator
of such processes.  The fraction of highly depleted stars in some
clusters, such as M67 (\cite{JFS99}) is significantly higher than
the norm, which could indicate at minimum that there is more than one
cause for strong lithium depletion.  Excluding stars that do not fit
an expected pattern also amounts to an effective prior on the sample
statistics.  If highly depleted stars are {\it a priori} excluded from
lithium samples then the highly depleted stars predicted by
theoretical models should also be removed when doing statistical
comparisons.  If we remove the observed upper limit and the upper 5 \% of
depletion factors from the theoretical models on the grounds that they
are rejected from samples of lithium abundances, we recover the same
(or higher) probabilities as we infer from including the one upper
limit in the sample.  Therefore, in this and subsequent tests we
retain the entire sample for statistical comparisons.

In constrast to the behavior around the sample median, the number of
outliers sets stronger constraints on stellar depletion.  The highest
depletion case of 0.5 dex is ruled out at the 95\% confidence level
even if we use the most generous definition of what constitutes an
outlier; this is a considerably stronger test than a similar
confidence level for a KS test.
  The Gaussian error model has severe difficulty reproducing
the observed number of outliers; it is excluded at the 90\% confidence
limit 0.05 dex below the median and at higher than a 99.9\% confidence
level 0.10 dex below the median.  Formally, a model with 0.13 dex 
depletion would
provide the best fit to the outlier fraction.  The 0.32 dex depletion
case is not formally excluded, but it is certainly disfavored by the
present data set.

We can set some rough bounds on stellar depletion based on the RNB
data sample without considering the effects of chemical evolution 
(discussed in Section 3) or possible systematic differences in
equivalent width measurements (see Section 4.)  Stellar depletion
at the 0.1 dex level is fully consistent with the data.  Models with
depletion as high as 0.5 dex are less than 5\%
probable, while models with no depletion are less than 10\% probable.  
PWSN compared the same models with the full
T94 data set and concluded that stellar depletion at the 0.2 dex level
provided the best fit to the dispersion in the T94 data set; a range
of 0.2 to 0.4 dex depletion was the result of several different
diagnostics of stellar depletion including the presence of highly
depleted stars and \6li to \7li ratio measurements and limits.  The
base RNB data set provides a lower central value for depletion; but 
because of the small sample size the bounds on depletion are actually 
widened relative to the conclusions of PWSN.  In the next sections we
consider other effects, and will return to our final estimate of the
primordial lithium abundance in section 5.

\section{Trends with Metal Abundance}

The dispersion in the RNB sample exceeds their quoted observational 
errors; as we have shown above it is consistent with the theoretical
predictions of mild rotational mixing.  However, RNB concluded that
the excess dispersion in their sample could be explained instead by
post-BBN galactic production of lithium.  As evidence for this they 
performed fits of lithium versus iron adopting for the functional 
form a fit which is a power-law in Li/H versus Fe/H: ~log(Li/H) = 
log(Li/H)$_P$ + a[Fe/H], where [Fe/H] $\equiv$ log(Fe/Fe$_{\odot}$).  

Although this form may provide a good fit to the data over a limited 
range in metallicity, it certainly cannot describe the evolution of 
an element whose BBN abundance is expected to provide the dominant 
contribution to its halo abundance.  To account for a significant 
BBN component along with a chemical evolution component that may 
scale linearly with the iron abundance (see for example
\cite{Ryan00}),
 the fitting function should be of the form Li/H = (Li/H)$_P$ 
+ b(Fe/Fe$_{\odot}$).  Since post-BBN, early galactic production of 
lithium may be dominated by cosmic ray nucleosynthesis which depends 
more on the oxygen than the iron abundance, \cite{Ryan00} also 
considered the consequences of an increasing oxygen abundance at 
low iron abundance.  In this case a linear fit to lithium 
as a function of oxygen would take the form
Li/H = (Li/H)$_{P}$ + b(Fe/Fe$_{\odot})^{0.7}$.
They found significant slopes 
ranging from $4.0 \times 10^{-9}$ to $1.8 \times 10^{-8}$ (in the 
{\it linear} iron $--$ {\it linear} lithium plane) and ranging from 
$0.9 \times 10^{-9}$ to $3.4 \times 10^{-9}$ (in the {\it linear}
oxygen $--$ {\it linear} lithium plane) on the assumption that the 
controversial
claims of very high oxygen abundance at low iron abundance are 
correct (\cite{IGR98}; \cite{Boe99}; but see also \cite{FK99}, 
 \cite{K00}).  Although we obtain  somewhat smaller 
slopes, we will show that the most important feature of these 
chemical evolution fits is that they do not explain the outliers
seen in the RNB sample.  Therefore, in contrast to RNB, we find 
that chemical evolution cannot account for the excess dispersion 
observed in their sample.

There are several issues which effect the quantitative fits for
the possible early (low-metallicity) evolution of lithium.  For 
example, the fits depend on the adopted stellar metallicities and 
RNB included two sets of metallicity estimates.  A literature 
value was taken from \cite{RN91}, \cite{RNB91}, \cite{CLLA}, 
and (for one star) \cite{Bee92}.  
There was also a 1 angstrom resolution estimate directly obtained 
by RNB for 21/22 detections in their sample.  Since no error 
estimates are quoted in the paper, we estimated them in two ways.  
The rms difference between the two sets is 0.14 dex, consistent 
with a $1 \sigma$ error of 0.1 dex in each.  This is also consistent 
with the error estimates in the primary sources used by RNB for 
the ``literature'' values.  Furthermore, there is a zero-point
difference of 0.13 dex between the literature and RNB metallicities,
in the sense that the RNB values are higher.  This different
metallicity zero-point contributes to the range in the inferred
chemical evolution slopes, in the sense that the slope inferred 
from the RNB metallicities is smaller than that obtained with the
literature metallicities.  Because the corrections to the lithium
abundances are small, only in the literature case does the error
in the metallicity have an impact on the overall dispersion (raising
$\sigma$ from 0.035 to 0.040 in the most extreme case.)

In addition, the RNB literature metallicities 
have an embedded effect that produces a significant component of 
the higher slope.  Two of the sources $--$ \cite{RN91}, \cite{RNB91}
 $--$ are systematically 0.15 dex lower than \cite{CLLA} because of 
a difference in the assumed solar iron abundance.  
The \cite{RNB91} abundances were corrected to the \cite{CLLA}
scale, but the \cite{RN91} values were not; RNB 
also did not use RN91 or Carney when there was an abundance from 
\cite{RNB91}.  To test for the importance of this effect we used the 
same primary sources, but corrected
\cite{RN91} to the \cite{CLLA} scale.  We then averaged multiple 
measurements weighted by their 
respective errors.  This reduces the
rms scatter compared with the RNB 1 A metallicities by 25 percent, 
and the slope also drops by 25 percent.  The direct $1\sigma$ error 
in [Fe/H] is 0.08 dex.  We therefore conclude that half of the
difference between the literature and RNB abundances is caused by 
the combination of data from different sources in the 
RNB literature values and the other half is the metallicity zero 
point.  We use the published literature RNB data for comparison with
other papers that have used this data; we believe that the homogeneous
RNB metallicities are a better choice for chemical evolution studies.

In Table 2 we show the data we used.  The abundance errors were 
estimated by adding in quadrature the $T_{\rm eff}$ error, the 
RNB slope of 0.065 dex per 100 K, and the RNB equivalent width 
error in the linear curve of growth approximation.  We obtain 
an average sample error of 0.036 dex rather than the RNB value
of 0.032 dex; we have not been able to trace the origin of the 
latter number in the RNB paper.  The T94 abundances have been
converted to the RNB temperature scale using the temperature
correction above; the T94 errors were estimated from the T94 equivalent
width errors and the temperature errors as described above.  The
average T94 error is 0.06 dex; we defer a discussion of the T94 data
to section 4.

\begin{deluxetable} {llllllll}

\tablecaption{Observational Data}
\tablehead{
\colhead{Star} & \colhead{Lit.} & \colhead{RNB} &
\colhead{$T_{eff}$} & \colhead{RNB} & \colhead{RNB}
& \colhead{T94} & \colhead{T94} \\
 & \colhead{[Fe/H]} & \colhead{[Fe/H]} &
\colhead{(K)} & \colhead{[Li]} & \colhead{EW(mA)}
& \colhead{[Li]} & \colhead{EW(mA)}
}
\tabletypesize{ \footnotesize}
\startdata
LP 651-4 & -2.96 & -2.60 & $6240 \pm 30$ & $2.11 \pm 0.039$ & $19.6 \pm 1.6$ & \nodata & \nodata \\
G4-37    & -2.73 & -2.70 & $6050 \pm 40$ & $2.11 \pm 0.045$ & $25.9 \pm 2.3$ & $2.08 \pm 0.083$ & $19 \pm 3.4$ \\
LP 831-7 & -3.25 & -3.32 & $6050 \pm 20$ & $2.07 \pm 0.029$ & $23.1 \pm 1.4$ & $2.18 \pm 0.051$ & $23 \pm 2.6$ \\
CD $-33  1173$  & -3.14 & -2.91 & $6250 \pm 20$ & $2.06 \pm 0.032$ & $17.2 \pm 1.2$ & $1.99 \pm 0.089$ & $12 \pm 2.4$ \\
BD +3 740 & -2.78 & -2.70 & $6240 \pm 40$ & $2.11 \pm 0.035$ & $19.5 \pm 1.1$ & $2.34 \pm 0.051$ & $24 \pm 2.4$ \\
BD +24 1676  & -2.71 & -2.38 & $6170 \pm 30$ & $2.10 \pm 0.047$ & $21.1 \pm 2.2$ & $2.37 \pm 0.049$ & $28 \pm 2.9$ \\
BD +20 2030 & -2.71 & -2.64 & $6200 \pm 40$ & $2.11 \pm 0.048$ & $20.5 \pm 2.0$ & $2.29 \pm 0.049$ & $23 \pm 2.2$ \\
BD +9 2190 & -2.89 & -2.83 & $6250 \pm 30$ & $2.00 \pm 0.042$ & $14.6 \pm 1.3$ & $2.19 \pm 0.085$ & $18 \pm 3.4$ \\
BD +1 2341p & -2.82 & -2.79 & $6260 \pm 40$ & $2.09 \pm 0.046$ & $17.8 \pm 1.6$ & $2.34 \pm 0.057$ & $21 \pm 2.6$ \\
HD 84937 & -2.30 & -2.12 & $6160 \pm 30$ & $2.17 \pm 0.027$ & $24.9 \pm 1.1$ & $2.25 \pm 0.053$ & $22 \pm 2.5$ \\
BD -13 3442 & -2.99 & -2.79 & $6210 \pm 30$ & $2.12 \pm 0.034$ & $21.0 \pm 1.4$ & $2.42 \pm 0.053$ & $30 \pm 3.4$ \\
G 64-12  & -3.17 & -3.24 & $6220 \pm 30$ & $2.14 \pm 0.029$ & $21.2 \pm 1.1$ & $2.38 \pm 0.059$ & $28 \pm 3.6$ \\
G 64-37  & -3.23 & -3.15 & $6240 \pm 30$ & $2.09 \pm 0.040$ & $18.2 \pm 1.5$ & $2.06 \pm 0.054$ & $14 \pm 1.6$ \\
BD +26 2651 & -2.88 & \nodata & $6150 \pm 40$ & $2.12 \pm 0.038$ & $22.5 \pm 1.5$ & $2.20 \pm 0.064$ & $20 \pm 2.9$ \\
CD -71 1234 & -2.50 & -2.60 & $6190 \pm 30$ & $2.20 \pm 0.025$ & $25.9 \pm 0.9$ & $2.36 \pm 0.051$ & $27 \pm 2.9$ \\
BD +26 3578 & -2.54 & -2.24 & $6150 \pm 40$ & $2.15 \pm 0.032$ & $24.6 \pm 1.1$ & \nodata & \nodata \\
LP 635-1 & -2.65 & -2.66 & $6270 \pm 30$ & $2.15 \pm 0.032$ & $20.2 \pm 1.2$ & $2.34 \pm 0.074$ & $24 \pm 3.9$ \\
LP 815-4 & -3.05 & -3.00 & $6340 \pm 30$ & $2.09 \pm 0.046$ & $16.1 \pm 1.6$ & $2.35 \pm 0.046$ & $22 \pm 2.1$ \\
CS 22943-095 & -2.55 & -2.20 & $6140 \pm 40$ & $2.12 \pm 0.035$ & $23.0 \pm 1.3$ & \nodata & \nodata \\
CD -35 14849 & -2.63 & -2.38 & $6060 \pm 20$ & $2.17 \pm 0.024$ & $28.8 \pm 1.4$ & \nodata & \nodata \\
G126-52  & -2.57 & -2.45 & $6210 \pm 40$ & $2.08 \pm 0.044$ & $19.1 \pm 1.6$ & $2.35 \pm 0.066$ & $26 \pm 3.6$ \\
CD -24 17504 & -3.55 & -3.24 & $6070 \pm 20$ & $1.97 \pm 0.033$ & $18.1 \pm 1.3$ & $2.15 \pm 0.072$ & $21 \pm 3.4$ \\
G186-26  & -2.85 & \nodata & 6180 & \nodata & \nodata &  $<1.36$ & $<2$ \\
\enddata
\end{deluxetable}

We considered two sets of metal abundances and two fitting functions
in $T_{\rm eff}$
(linear and power law), for four basic cases.  As already noted by RNB,
much of the slope comes from three outliers in the sample; we
therefore repeated the analysis for all four sets with the same outliers 
excluded as discussed in RNB.  In Table 3 we present the eight sets of 
results.  The
cases are identified in column 1; the first four include outliers and
the last four do not.  The cases starting with L use the RNB [Fe/H]
metal abundances which yield a low slope; the cases starting with H
use the literature [Fe/H] values which yield a high slope.  The cases
ending with L are linear fits and the cases ending with P are power
law fits.  The zero-point and slope of the different fits are in columns 
2 and 3.  The median abundance corrected to zero metal abundance and both 
the predicted and actual residual dispersion are given in columns 4 to 6.  
The different cases are illustrated in Figures 2 and 3.

\begin{deluxetable} {llllll}
\tablecaption{Chemical Evolution Fits}
\tablehead{
\colhead{Case} & \colhead{$10^{10} Li_p$} & \colhead{$10^{10}$ slope} &
\colhead{Median [Li]} & \colhead{$\sigma$(pred)} & \colhead{$\sigma$(obs)}
}
\startdata
HL & $1.11 \pm 0.05$ & $97.5 \pm 24.7$ & 2.045 & 0.041 & 0.046 \\
LL & $1.17 \pm 0.05$ & $44.3 \pm 16.6$ & 2.065 & 0.039 & 0.049 \\
HP & $1.04 \pm 0.06$ & $21.4 \pm 5.0$ & 2.015 & 0.044 & 0.048 \\
LP & $1.12 \pm 0.06$ & $11.4 \pm 4.0$ & 2.05 & 0.039 & 0.051 \\
HLno & $1.21 \pm 0.03$ & $56.8 \pm 17.0$ & 2.070 & 0.038 & 0.045 \\
LLno & $1.22 \pm 0.03$ & $29.3 \pm 9.8$ & 2.080 & 0.037 & 0.049 \\
HPno & $1.16 \pm 0.04$ & $12.8 \pm 3.6$ & 2.055 & 0.039 & 0.046 \\
LPno & $1.19 \pm 0.04$ & $7.4 \pm 2.4$ & 2.065 & 0.038 & 0.050 \\
\enddata
\end{deluxetable}

There are two important conclusions to be drawn from this exercise.  
First, the chemical evolution slopes are sensitive to {\it all} of 
the assumptions in the models, with a wide range of slopes possible.  
Second, detrending the data in the linear Li versus Fe plane does not 
bring the outliers onto the mean trend.  In all cases the formal 
dispersion of the 
detrended samples are larger than the estimated errors. 
This can be traced directly to the presence 
of outliers whose lithium abundance differs significantly from the 
sample mean.  Intuitively this can be easily understood; because 
the absolute metal abundances of the stars are small, there is 
little room for a significant chemical evolution correction.  There 
are three stars noted as outliers in the RNB chemical evolution
analysis: CD -24 17504 ([Li]=1.97 $\pm$ 0.033, $4.2 \sigma$ below 
the mean); BD +9 2190 ([Li]=2.0 $\pm$ 0.042, $2.6 \sigma$ below 
the mean); and CD -71 1234 ([Li]=2.20 $\pm$ 0.025, $3.6 \sigma$ 
above the mean).  In the linear fit to the literature iron abundances 
these three stars are respectively $2.4 \sigma$ below, $2.3 \sigma$ 
below, and $2.0 \sigma$ above the mean; for the linear fit to the 
RNB iron abundances the same stars are respectively $3.3 \sigma$ 
below, $2.3 \sigma$ below, and $3.8 \sigma$ above the mean.  There 
is also the upper limit in G186-26, making a total of 4/23 outliers 
{\it regardless of the presence or absence of chemical evolution 
detrending.}  Similar results apply to the power-law fits.

We have not performed chemical evolution fits 
for our rotationally mixed models because it is not a well-posed
problem for such a small sample of lithium abundances; a 
metallicity-dependent distribution 
of stellar depletion factors needs to be convolved with a mean 
chemical evolution trend.  From PWSN, we can anticipate that 
the contribution of a range of metallicities to the dispersion
will be small and difficult to detect in a sample of this size.

\subsection{Comparison of Theory and Observation Including Chemical
Evolution Detrending}

Table 3 presents the results of KS and outlier test comparisons of 
the models and data under different chemical evolution detrending 
scenarios.  We have used the same theoretical models as in section 2.
As noted in Table 3, the additional observational errors from the
uncertainty in lithium production does not significantly impact the
overall observed error.   
Because the various fits yield similar conclusions, in Figure 4 we
show only the most probable of these cases.  The
observational data in in Figure 4 is the cumulative distribution of 
[Li] from the linear fit to the RNB metal abundances, corrected to
zero metal abundance.  We compare this data set to a gaussian with
$\sigma = 0.04$ dex and the same three 
theoretical distributions as in Figure 1.  The qualitative trends 
are similar to those obtained with the base RNB data.

\begin{deluxetable} {lllllll}
\tablecaption{Chemical Evolution Probabilities}
\tablehead{
\colhead{Case} & \colhead{Outlier Test} & & & \colhead{KS test} & & \\
& \colhead{s0} & \colhead{s0.3} & \colhead{s1} & 
\colhead{s0} & \colhead{s0.3} & \colhead{s1}
}
\startdata
None & 0.45 & 0.15 & 0.05 & 0.49 & 0.13 & 0.06 \\
LL & 0.31 & 0.11 & 0.04 & 0.32 & 0.15 & 0.05 \\
LP & 0.38 & 0.15 & 0.06 & 0.52 & 0.18 & 0.04 \\
HL & 0.38 & 0.13 & 0.045 & 0.34 & 0.14 & 0.03 \\
HP & 0.35 & 0.12 & 0.04 & 0.66 & 0.19 & 0.06 \\
\enddata
\end{deluxetable}

The five cases considered are
no chemical evolution; low slope, linear Li-Fe (LL),
low slope, power law Li-Fe (LP), high slope linear (HL),
high slope power law (HP).  The first three columns are the 
probabilities of drawing the data from the theoretical s0, s0.3,
and s1 distributions.  The no evolution case is evaluated at
0.1 dex below the median; the other cases are smoother and are
evaluated in 0.01 dex increments between 0.05 and 0.1 dex below the
median and averaged.  The second set of three columns are the
KS test probabilities for the same cases and theoretical
distributions.

In Table 5, we give both the zero points and inferred primordial
abundances for the different cases on the RNB abundance scale; we
argue elsewhere that these should be adjusted up by 0.1 dex because of
systematic model atmosphere/temperature scale effects.

\begin{deluxetable} {lllll}
\tablecaption{Inferred Primordial Lithium}
\tablehead{
\colhead{Case} & \colhead{Observed} & \colhead{s0} & \colhead{s0.3} & 
\colhead{s1}
}
\startdata
None & 2.11 & 2.29 & 2.43 & 2.61 \\
LL & 2.07 & 2.25 & 2.39 & 2.57 \\
LP & 2.05 & 2.23 & 2.37 & 2.55 \\
HL & 2.05 & 2.23 & 2.37 & 2.55 \\
HP & 2.02 & 2.20 & 2.34 & 2.52 \\
\enddata
\end{deluxetable}

\subsection{Chemical Evolution Implications for $^{6}$Li}

If, indeed, the abundance of lithium is evolving at very low-metallicity
as RNB suggest, the most likely source of post-BBN lithium is from cosmic
ray nucleosynthesis (\cite{rfh}).  One consequence of CRN is the concommitant 
production of \6li along with \7li resulting in comparable amounts of 
post-BBN production of both isotopes.  At very low-metallicity the 
lithium isotope production is dominated by $\alpha - \alpha$ fusion
(Steigman and Walker 1992) leading to a 7/6 production ratio of $R_{76} 
\approx 1.6$ (Kneller, Phillips, and Walker 2000).  At higher metallicity 
this ratio decreases slightly to $R_{76} \approx 1.5$ (Steigman and Walker 
1992; Kneller, Phillips, and Walker 2000). 
As a result, the \6li/\7li ratio provides a means to test the RNB hypothesis
that the lithium abundance is increasing at a noticeable rate in the early
Galaxy at very low metallicity ([Fe/H]$~\la -2$).  If the observed lithium
abundances (without allowance for depletion by rotational mixing) are fit 
to a metallicity relation of the form Li/H $= a + bx$, where $a \equiv 
($Li/H)$_{P}$ and $x$ is either Fe/Fe$_{\odot}$ or (Fe/Fe$_{\odot})^{0.7}$, 
the predicted 7/6 ratio is

\begin{equation}
^{7}Li/^{6}Li = R_{76} + {(1 + R_{76})a \over bx}  
\end{equation}

At present, detections of \6li are claimed for three metal-poor stars 
(Smith, Lambert, \& Nissen 1993, 1998, Hobbs, \& Thorburn 1991, 1997, 
Hobbs, Thorburn, \& Rebull 1999, Nissen \etal 1999,  Nissen \etal 2000) 
two of which are included in the RNB sample.  In Figure 5 we compare 
the observed 6/7 ratios with those predicted by RNB evolution for $1.5 
~\la R_{76} ~\la 1.6$.    While the post-BBN evolution suggested by RNB 
may account for the observed 6/7 ratio in one (possibly two) stars, it 
is clear that it is too rapid to satisfy all the observational data.

Of course, if we are correct that some depletion via rotational mixing 
can have occurred in one or more of these stars, it may be that the 
observed 6/7 ratios are not representative of the prestellar values.  
As an illustration, we show by the open circles in Figure 5 the 
predictions for our standard (\ie, no rotational mixing or gravitational
settling) model depletion .  We emphasize though that these ``predicted'' 
data points should not be compared to the evolution curves which have 
been derived on the assumption of no depletion.  Nonetheless, it is 
clear that no simple model for post-BBN lithium production can account 
for all three data points in the absence of some \6li depletion.  We 
note that the two stars in the RNB sample (HD 84937 and BD +26 3578) 
have lithium abundances slightly above the (undepleted) plateau (2.17 
and 2.15 respectively); this may indicate that they are detectable 
because they are a bit underdepleted within the rotational mixing context.
 
\section{Observational Data}

In the preceding sections we have compared our results with the RNB
data sample reaching qualitatively similar conclusions to those drawn 
from earlier data sets, in particular from the large T94 sample that
we used in PWSN.  
However, there are some observational differences, and it is important 
to identify how and why the various observational data sets differ.  
RNB obtained lithium observations for 23 halo stars; one star (G186-26) 
had only an upper limit and was excluded from their dispersion anaylsis.  
The RNB sample of stars were chosen in a narrow $T_{\rm eff}$ range with 
a low metal abundance.  18 of the 22 remaining stars were also studied 
in T94, a sample designed with similar goals.  There were also 18 stars 
in common with the earlier \cite{Ryan96} data set and 10 stars in common 
with the \cite{Sp93} sample (\cite{Sp96}, \cite{Spites}, \cite{Sp84};
hereafter SS).  Because the \cite{Ryan96} sample is dominated by the 
large number of stars from the T94 sample, there are really only two 
independent, primary samples which may be compared with RNB: T94 and 
the SS sample.

\subsection{Comparison with Other Data Sets}

\subsubsection{Comparison with T94: The Origin of Differences in
Zero-point and Dispersion}

The conclusions drawn by RNB and T94 are markedly different despite
the significant overlap in the two samples and their similar goals
and design.  We therefore begin by examining the ingredients that 
could be responsible for this difference, namely 1) the statistics 
of the subset of the T94 data set studied by RNB, relative to the 
statistics of the entire T94 data set; 2) the choice of effective 
temperature scale; 3) the equivalent width measurements and; 4) 
the model atmospheres used to relate equivalent width and effective 
temperature to abundance.  

The raw dispersion measured by RNB for their sample of 22 stars 
(0.052 dex) is very similar to the raw dispersion for the subset 
of their 18 stars in common with T94 (0.054 dex), suggesting that 
the stars not in common do not strongly influence the overall result.  
The raw dispersion for the full T94 sample was 0.13 dex, similar 
to the dispersion of 0.12 dex that would be inferred for the subset 
of 18 stars from the T94 data.  Therefore, the RNB data set appears 
to be a fair subsample of the T94 data set.  This is not surprising 
since both were chosen using similar kinematic, metal abundance, 
and effective temperature criteria.  However, the average abundances 
for the stars in common derived by RNB and by T94 differ significantly, 
2.11 and 2.29 respectively.  

In subsequent steps we examined the impact of changes in the temperature 
scale and equivalent widths.  The $T_{\rm eff}$ scale chosen by RNB is 
different (and on average cooler) than that used by T94.  To estimate 
the importance of this effect, we compared the temperatures used by RNB 
and T94.  We used the RNB slope of 0.065 dex per 100K to infer the lithium 
abundances that T94 would have obtained using the RNB temperature scale.  
If the different temperatures adopted by T94 and RNB were responsible 
for the different conclusions about the sample dispersion we would expect 
a large decrease in the sample dispersion by performing this operation 
while retaining the T94 equivalent widths and model atmospheres.  Adopting 
the RNB temperatures reduces the average abundance inferred using the T94 
equivalent widths and atmosphere model only from 2.29 to 2.25, while actually 
slightly increasing the dispersion that would have been inferred from the 
T94 data relative to the T94 $T_{\rm eff}$ scale. The abundances that would
have been inferred from the T94 equivalent widths and model atmospheres 
with the RNB $T_{\rm eff}$ scale are given in Table 2 (see section 2).  
We conclude that while the choice of temperature scale does influence 
the abundance zero-point, the different temperature scales do not explain 
the difference in the dispersion of the samples.  We illustrate this in 
Figure 6, where the RNB abundances are compared with the T94 abundances 
for stars in common shifted to the same $T_{\rm eff}$ scale.  The intrinsic 
scatter is clearly larger for the T94 equivalent widths, even accounting 
for the larger formal equivalent width error bars.

We also derived the abundances that T94 would have obtained had both the 
lithium equivalent widths and temperatures of RNB been used instead of 
the EW and $T_{\rm eff}$ as adopted by T94.  The sole remaining difference 
after this has been done is the choice of model atmospheres relating 
equivalent width, temperature, and abundance.  For this test we used 
the linear curve of growth approximation; e.g. the corrected [Li/H] = 
[Li/H](T94) + log (EW RNB/EW T94).  Changing the equivalent widths 
(along with $T_{\rm eff}$) leads to a large decrease in the dispersion, 
from 0.13 dex to 0.07 dex; furthermore, the average inferred abundance 
decreases to 2.22.  The average difference between the equivalent width 
measurements of RNB and T94 is 1.9 mA (in the sense that T94 is 
systematically higher), so there is both a zero-point shift and a 
difference in the range of equivalents widths at fixed $T_{\rm eff}$ 
in the RNB sample relative to the T94 sample.

We attribute 
the remaining zero-point shift to one of two effects.  The linear curve 
of growth assumption that we have employed could introduce some errors; 
T94 and RNB also used different model atmospheres to relate abundance to 
equivalent width.  \cite{Ryan00} estimate the systematic differences 
arising from the different model atmospheres to be at the $\sim$ 0.08 dex 
level, which can account for all but a small (0.03 dex) difference in the 
mean abundance.  We therefore conclude that the reason for the significantly 
different RNB dispersion estimates from those of T94 are due to differences 
in the underlying basic equivalent width data and not by the choice of 
$T_{\rm eff}$, the sample properties, or the model atmospheres.  In contrast, 
the difference in the lithium abundance zero-point can be attributed to 
a combination of a different $T_{\rm eff}$ scale, systematically lower 
RNB equivalent widths relative to T94, and the choice of different model 
atmospheres.  This leads to a large overall difference between the mean 
abundances derived for stars in common, corresponding to a change in the 
inferred primordial lithium abundance comparable to the lower end of the 
range of stellar depletion presented in PWSN.

There is an average zero-point offset of 2.5 mA in the RNB sample relative 
to the SS data set; a similar effect compared was discussed in \cite{Ryan96}.  
The overall morphology is similar to that between T94 and RNB: 2 out of 10 
points differ by more than 2$\sigma$ even when the zero-point offset is 
taken into account.  For completeness, we note that there is also a 
zero-point offset of 0.8 mA relative to the \cite{Ryan96} analysis; 
as mentioned above, because this sample is heavily weighted by the T94 
sample we did not perform a separate comparison of the \cite{Ryan96} 
and RNB samples.  We therefore conclude that zero-point differences in 
equivalent width measurements appear to be significant, and by themselves 
they contribute an uncertainty of order 10 \% to the absolute abundances.

\subsection{Interpretation of the Differences}

Even before reducing the dispersion by appealing to chemical evolution 
trends, the RNB sample has a smaller dispersion than does the T94 sample.  
RNB attributed the differences between their data and that of T94 to an 
underestimate of the errors in the T94 data.  In particular, T94 did not 
correct for scattered light and sky subtraction.  RNB note that this could
increase the formal errors of individual data points in the T94 sample
by a factor up to 1.7.  In this case, one would not expect a gaussian 
distribution of the differences in equivalent widths, since the T94 stars 
with the largest relative errors from these effects would be affected more 
than those where the quoted T94 error estimates are accurate.  There is an 
independent test of this hypothesis: we can compare the results of SS with 
those of T94.  If the most discrepant points are due to larger than expected 
errors in the T94 measurements, then there is no reason to expect the SS
sample to have encountered the same problems.

In Figure 7 we compare equivalent width measurements for stars from different 
sources (RNB, T94, SS).  The three left-hand panels compare the eight stars 
with measurements from all three sources; the right panels compare RNB with 
the stars in common with the T94 and SS data sets, respectively.  Although 
the overlap among the samples is small (eight stars), we see no direct 
evidence that the T94 data is in conflict with the two other data sets; 
similarly, \cite{Ryan96} found a good correlation between the SS data 
and the T94 data once a zero-point offset was taken into account.

In light of the ambiguous results above, it is worth returning to the
question of what degree of stellar depletion is consistent with the
T94 data.  Because the RNB $T_{\rm eff}$ errors are significantly
smaller than the T94 errors, the average error using the T94
equivalent widths is reduced to 0.06 dex.  This provides a smaller 
but more precise sample than the one used in PWSN.
  
Adopting the T94 equivalent widths instead of the RNB equivalent
widths produces a significantly different cumulative distribution.
We compare the observed distribution with theoretical simulations
convolved with a $\sigma = 0.06$ dex gaussian in Figure 8.  The s0.3
case with 0.32 dex depletion is now the best fit, while the 0.18 and 
0.5 dex depletion cases are only marginally consistent.  We include 
this to emphasize that the differences between the observational data 
sets needs to be reconciled in order to set more precise bounds on 
stellar depletion.

In conclusion, we find that the large difference between the results 
of RNB and T94 in their dispersion analyses can be traced directly to 
equivalent width measurements.  The overall deviations exceed those 
predicted from the quoted errors.  There are significant zero-point 
offsets, and external comparison with a small overlap sample from SS 
does not clearly identify a problem with the T94 values.  We therefore 
caution that further observational work is likely needed to uncover the 
origin of the differences, particularly since the overall conclusions 
depend sensitively on the presence or absence of a small number of outliers.

\section{Discussion}

Knowledge of the primordial lithium abundance sets interesting constraints 
on Big Bang Nucleosynthesis.  However, the determination of the primordial
lithium abundance relies both on observational data as well as on the 
model for stellar depletion.  The most recent \7li abundance data sets 
exhibit a core with little internal scatter and a small number of outliers; 
these properties have been used to argue that there is little, if any, 
room for any stellar depletion.  We have analyzed the RNB data set, and 
find that with or without accounting for a trend with metal abundance the 
data is consistent with mild stellar depletion; the best fit depletion is 
in the range of 0.1 $--$ 0.2 dex .  Theoretical models with rotational mixing 
depletion this low predict a core with small scatter, since the large 
majority of young stars have low, and similar, rotation rates.  Therefore, 
the number of outliers is a stronger test of the presence or absence of 
dispersion from rotational mixing.  Either the no depletion case or models 
with lithium as depleted as 0.5 dex are unlikely based on both KS tests 
and the predicted number of overdepleted outliers as compared with the 
observed number.  Our results differ from those of RNB because they 
detrended the data in the log(Li) $--$ Log(Fe) plane rather than in the 
linear Li $--$ linear Fe plane which is more appropriate when testing for 
the presence of post-BBN \7li production.  Similar conclusions can be 
derived from the observed ratio of \6li to \7li.

We have also compared the T94 and RNB data sets, and find that the
different conclusions that the two papers drew about the dispersion in
lithium among halo stars can be traced directly to differences in
equivalent width.  If the T94 equivalent widths are used instead of
the RNB equivalent widths, a stellar depletion factor of 0.3 dex is
inferred.  We find no compelling evidence of problems in the T94 data
by comparing both it and RNB with an (admittedly small) set of stars
in common with other studies.  This indicates that there is further
work to do on the observational front before making sweeping claims
with implications for cosmology.  Systematic observational errors from
the temperature scale (0.05 dex), choice of model atmospheres (0.08
dex), and equivalent width zero-point errors (0.05 dex) alone yield an
uncertainty of 0.11 dex in the observed \7li abundance before
considering stellar depletion $--$ an error of similar magnitude to the
theoretical uncertainties.  In the last two sections we discuss 
the issues and uncertainties involved in the stellar modelling and the
implications for BBN.

\subsection{Stellar Physics Models}

Further improvements are also desirable in the theoretical modelling
of mixing and diffusion processes in the envelopes of low mass stars.
These fall into the general categories of improved stellar physics on
the one hand and better knowledge of 
the distribution of initial conditions and the angular momentum loss
law on the other hand.

In Population I stars we have extensive empirical data on the
distribution of rotation rates as a function of mass and age.  We can
only observe very metal poor stars when they are old, and therefore
must extrapolate the behavior of Population I stars into a different
metallicity regime.  The best prospect for constraining the theory is
in observations of young clusters of intermediate metallicity; this
will permit a direct test of the distribution of rotation rates and
their time evolution.  If the fraction of rapid rotators is different
from that present in Population I open clusters, the predicted number
of outliers would be affected.  More efficient angular momentum loss
in metal-poor stars could also reduce the predicted number of
overdepleted stars for a given absolute depletion.

On the stellar physics side, the important uncertainties are internal
angular momentum transport and the interaction of gravitational
settling and rotational mixing.  Helioseismic data indicates that the
solar internal rotation is independent of depth in the radiative core
down to 0.2 solar radii (e.g. \cite{Chap99}); the situation in deeper
layers is less certain (compare \cite{Chap99} with \cite{Gav00}).
The spindown of young open cluster stars, however, is not consistent
with uniform rotation enforced on a very short timescale
(\cite{KPBS}).  This combination implies that the timescale for
effective angular momentum coupling between the surface and interior
is intermediate between the ages of the young open clusters (50-100
Myr) and the Sun (4.57 Gyr.)  We are currently evaluating models in
the limiting case of uniform rotation at all times to infer the impact
on the predicted depletion.  The general sense would be to reduce
lithium depletion in models with shallower convection zones (because
the diffusion coefficients are larger if the core rotates more rapidly
than the surface.)  Therefore the predicted degree of depletion in
halo stars for a given solar calibration will be reduced.  However, a
range of stellar depletion factors for a range in solar initial
conditions will still be possible; the net effect will be to make the
observed halo star depletion consistent with less extreme values of
the solar initial conditions.

Finally, gravitational settling and microscopic diffusion could affect
surface lithium abundances.  The gravitational settling of helium will
produce a mean molecular weight gradient below the surface convection
zone; composition gradients could reduce the effect of mixing (see
\cite{Z92} for a discussion.)  If rotational mixing was simply
suppressed, however, models with gravitational settling predict a
decrease in halo star surface lithium with increased effective 
temperature which is not observed (\cite{Cha92}).  \cite{Vau99} has
raised the possibility of a nonlinear interaction that results in the
suppression of both mixing and diffusion.  This is an interesting
possibility that should be investigated.  There are, however, some
factors that make a complete cancellation unlikely in our view.

First, the physical conditions in Population I stars with temperatures
similar to the plateau stars are not very different from the halo star
conditions.  We observe strong depletion and dispersion in M67 stars
with temperatures around 6200K, which suggests that mixing is not
inhibited in solar abundance stars where the timescale for
gravitational settling is similar to the halo star case.  In addition,
\cite{CDP95} investigated the interaction of gravitational settling
and mixing for an earlier generation of models.  They found that the
time and mass dependence of lithium depletion was difficult to
reconcile with a strong suppression of mixing by settling.  It is
important to test the interaction of these physical processes against
Population I data.  In addition to the lithium question, such models
must address the apparent absence of a gravitational settling
signature in the turnoff region of globular clusters
(\cite{Cha92}; \cite{BV01}).

\subsection{Implications for BBN}

Uncertainties in lithium equivalent width measurements, temperature
scales, and model atmospheres introduce a {\it systematic} error in
the determination of lithium abundances (on the log scale) which we
estimate as $\pm 0.1$~dex, in agreement with \cite{Ryan00}'s detailed 
analysis of the error budget.   Reflecting this uncertainty, the level 
of the Spite plateau, before accounting for depletion by rotational 
mixing or for post-BBN lithium production, has been variously estimated 
as 2.1 (RNB), 2.2 (\cite{BM97}, \cite{BMP97}), and 2.3 (T94).  We adopt 
2.2 $\pm~0.1$ for our baseline estimate.  We have found that the residual 
dispersion in the RNB data is well accounted for in a model of stellar 
depletion induced by rotational mixing, even without account of the 
additional dispersion that might be due to a real spread in halo star 
abundances due to post-BBN lithium production.  Our best estimate of 
the overall depletion factor consistent with the RNB data set is 0.13 
dex, with a 95\% range extending from 0.0 to 0.5 dex.  Similarly, using 
the T94 equivalent widths we find an overall best-fit depletion of 0.32 
dex.  For all of the reasons given above, we believe that modest stellar
depletion factors are consistent with the data.  At the same time, the
sample size is small and these conclusions are subject to Poisson
noise.  Given that the expected fraction of overdepleted stars from
rotational mixing is of order 15\%, even adding or subtracting a
single star from the sample can have a significant impact on the
inferred depletion.  With this in mind, we adopt an overall depletion 
factor of 0.2 $\pm~0.1$ dex and, adding this correction 
to our baseline estimate (and combining these systematic uncertainties 
linearly), we derive a primordial lithium abundance of 2.4 $\pm~0.2$.

Having established an observed halo lithium abundance of $2.2\pm 0.1$ and 
our best estimate of the primordial lithium abundance of $ 2.4 \pm 0.2$
(based on both the observed abundance and a theoretical determination
of the depletion required by the dispersion in the observed abundance), we
can now compare to the primordial lithium abundance predicted by standard
BBN.   As previously mentioned, in standard BBN the predicted abundances 
of the light nuclides is a function of one parameter, the baryon-to-photon 
ratio $\eta$.  Within the errors introduced by uncertainties in the weak 
and nuclear cross sections, the predicted abundances of D, \3he, \4he, 
and \7li follow from a determination of $\eta$.  The predicted abundance
of deuterium depends most strongly on $\eta$ and, using the observed D/H
as measured in high-redshift QSO absorption line systems (Burles \& Tytler
1998a,b; O'Meara \etal 2000), it can be used as a ``baryometer'': $\eta_{10} 
\equiv 10^{10}(n_{B}/n_{\gamma})= 5.6 \pm 0.5$ ($1\sigma)$.  The predicted 
lithium abundance corresponding to this range of baryon densities is [Li]$_{P} 
= 2.3$ to 2.8 ((Li/H)$_{P} = 2 - 6 \times 10^{-10})$, inconsistent with the 
Ryan \etal 2000 halo star lithium abundance, but in good agreement with our 
estimate for the primordial lithium abundance.  Our range of predicted Li/H 
is based on the OSW 2000 $1\sigma$ range (\cite{OSW} and references therein), 
which is broader than, but consistent with that of \cite{BNT}, which for the 
same $\eta$ range predicts [Li]$_{P} = 2.4$ to 2.7.  The NACRE collaboration's 
(\cite{NACRE}) compilation yields a similar but slightly lower range 
than that of \cite{BNT}.  We note that recent deuterium observations 
(\cite{D'Od01}, \cite{Pettini}) suggest that the damped Lyman 
$\alpha$ systems may have systematically lower D/H than the Lyman limit 
systems.  The lower D/H would correspond to an even higher BBN Li/H, 
further exacerbating the disagreement with the Ryan \etal 2000 lithium 
abundance, and even pushing the upper envelope of our model-dependent 
estimate of the primordial abundance.  The situation is summarized in 
Figure 9, where we show as a band, the standard BBN-predicted abundance 
of Li/H as a function of the BBN-predicted D/H. The ``data point" is 
for the O'Meara \etal 2000 deuterium abundance and the Ryan \etal 2000 
lithium value.  The horizontal band corresponds to our estimate of the 
depletion-corrected primordial lithium abundance. The Ryan \etal 2000 
inferred primordial lithium abundance is too small, by a factor or two 
or more, to be consistent with the deuterium-based BBN prediction.  In 
contrast, there is excellent overlap between our depletion-corrected 
estimate ([Li]$_{P} = 2.4 \pm 0.2$) and the deuterium-constrained BBN 
prediction.  Consistency with BBN {\it requires} that lithium has been 
depleted in the metal-poor halo stars of the Spite plateau by an amount 
consistent with that predicted by our models of rotational mixing induced 
stellar depletion.  

\acknowledgements

M.P. wishes to acknowledge support from the NASA Astrophysics Theory
Program (grant NAG5-7150.)  The work of G.S. and T.P.W. is supported
by DOE grant DE-FG02-91ER40690.

\figcaption{
The cumulative observed distribution of RNB is compared with the
distributions expected from Gaussian errors with $\sigma$ = 0.035 dex
(short dashed line) and the s0 (medium-dashed line), s0.3 (dot-dash line), and
s1 (long-dashed line) distributions from PWSN convolved with the
same error.  The zero-point of
the theoretical distributions is set by anchoring the median depletion
factors of 0.18 dex, 0.32 dex, and 0.50 dex respectively at the sample
median of 2.11.
}

\figcaption{
Chemical evolution fits to the RNB data set using
the RNB literature metal abundances (top panel) and the RNB low
resolution metal abundances (bottom panel) in the linear Li - linear 
Fe plane.  The solid line includes all stars; the dashed lines
excludes the stars identified as outliers in RNB.
}

\figcaption{
Chemical evolution fits to the RNB data set using
the RNB literature metal abundances (top panel) and the RNB low
resolution metal abundances (bottom panel) under the assumption of
strong oxygen
enhancement in metal poor stars.  We mapped Fe onto Z
as described in the text and performed a
least-squares fit in the linear Li - linear 
Z plane.  The solid line includes all stars; the dashed line
excludes the stars identified as outliers in RNB.
}

\figcaption{
The RNB data presented here has been corrected to zero metal abundance
using the LL model slope including all stars.
The cumulative observed distribution of RNB corrected for chemical
evolution is compared with the
distributions expected from Gaussian errors with $\sigma$ = 0.04 dex
(short dashed line) and the s0 (medium-dashed line), s0.3 (dot-dash
line), and s1 (long-dashed line) distributions from PWSN.  The zero-point of
the theoretical distributions is set by anchoring the median depletion
factors of 0.18 dex, 0.32 dex, and 0.50 dex respectively at the sample
median of 2.07.
}

\figcaption{
The measurements of \6li to \7li (filled circles) are compared with the
predictions of the different chemical evolution models; the solid
lines correspond to the range of uncertainties for the high-slope
cases and the dashed lines correspond to the range of uncertainties
for the low slope cases.  The top panel uses the linear Li - linear Fe
fits and the bottom panel estimates the effect of strong oxygen
enhancement in the most metal-poor stars in the same fashion as Figure
3.  See text for data sources; the open circles include a
classical model \6li depletion factor of 0.2 dex for the two more
metal-rich detections.
}

\figcaption{
The RNB data set (top panel) is compared with the T94 data set under
three different assumptions.  The published T94 data for stars in
common is presented in the second panel.  The third panel shows the
effect of replacing the T94 $T_{\rm eff}$ values with those from RNB.
The bottom panel shows the effect of replacing both the temperatures
and equivalent width measurements of T94 with the RNB values under the
assumption of a linear curve of growth.}

\figcaption{
Equivalent width measurments from RNB, T94, and SS are
compared in this figure.  The left three panels include stars in
common to all three sets.  RNB data are compared to SS data in the top
left panel; RNB compared to T94 data in the middle left panel; and SS
compared to T94 in the bottom left panel.  The two right panels
compare the RNB data with the full overlap sample of SS (top right)
and T94 (bottom right.)}

\figcaption{
The cumulative observed distribution of the RNB sample using the T94
model atmospheres and equivalent widths is compared with the
distributions expected from Gaussian errors with $\sigma$ = 0.06 dex
(short dashed line) and the s0 (medium-dashed line), s0.3 (dot-dash
line), and s1 (long-dashed line) distributions from PWSN 
convolved with the same error.  The zero-point of
the theoretical distributions is set by anchoring the median depletion
factors of 0.18 dex, 0.32 dex, and 0.50 dex respectively at the sample
median of 2.25.}

\figcaption{
The solid curves show the $\pm 1\sigma$ range for the BBN-predicted relation
between primordial lithium and primordial deuterium.  The point with error
bars is the O'Meara \etal deuterium value and the Ryan \etal (2000) lithium
estimate.  The horizontal band is our depletion-corrected lithium estimate,
and the vertical band indicates the observed range for deuterium.}

\begin{figure}
\plotone{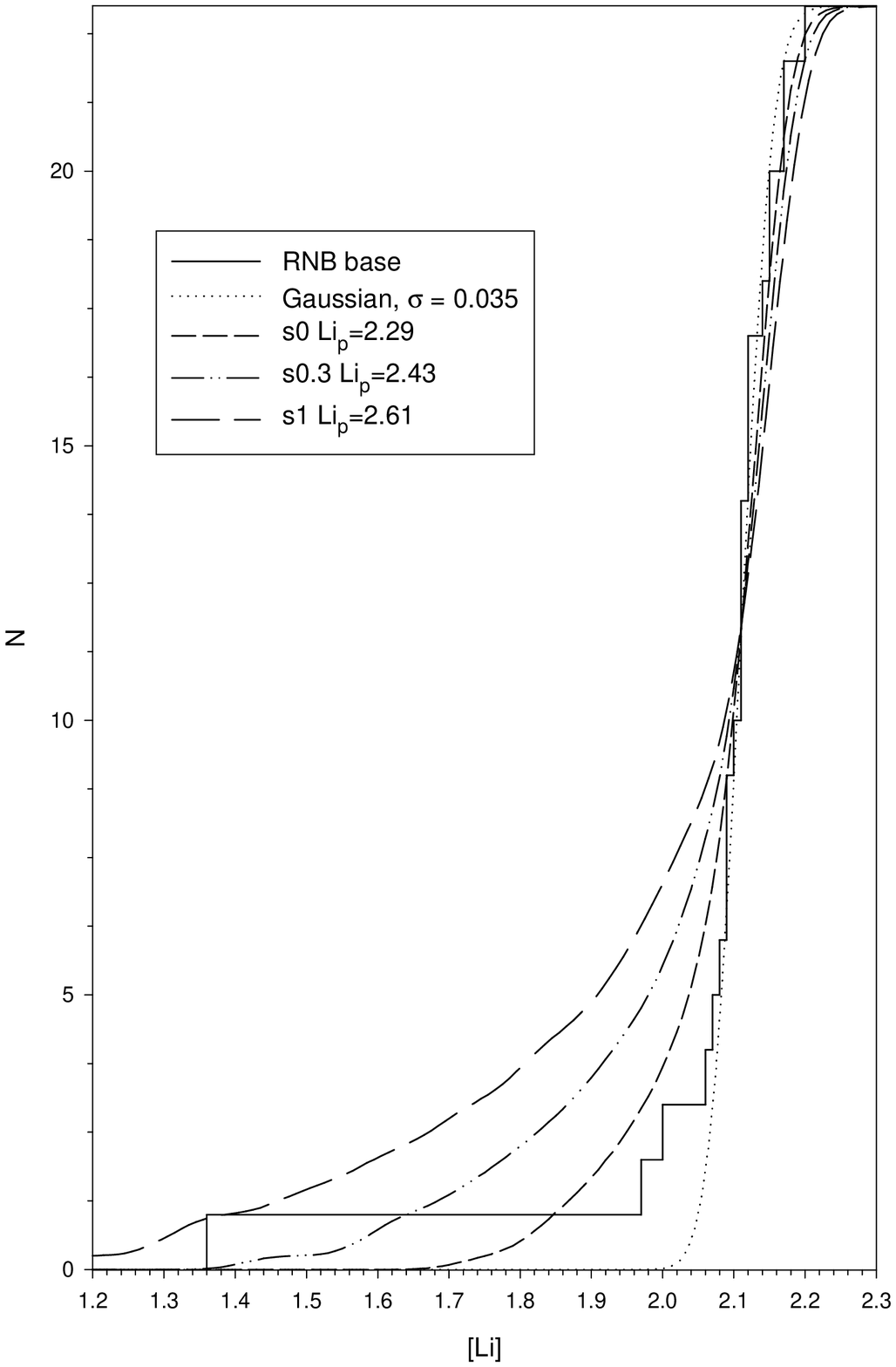}
\end{figure}

\begin{figure}
\plotone{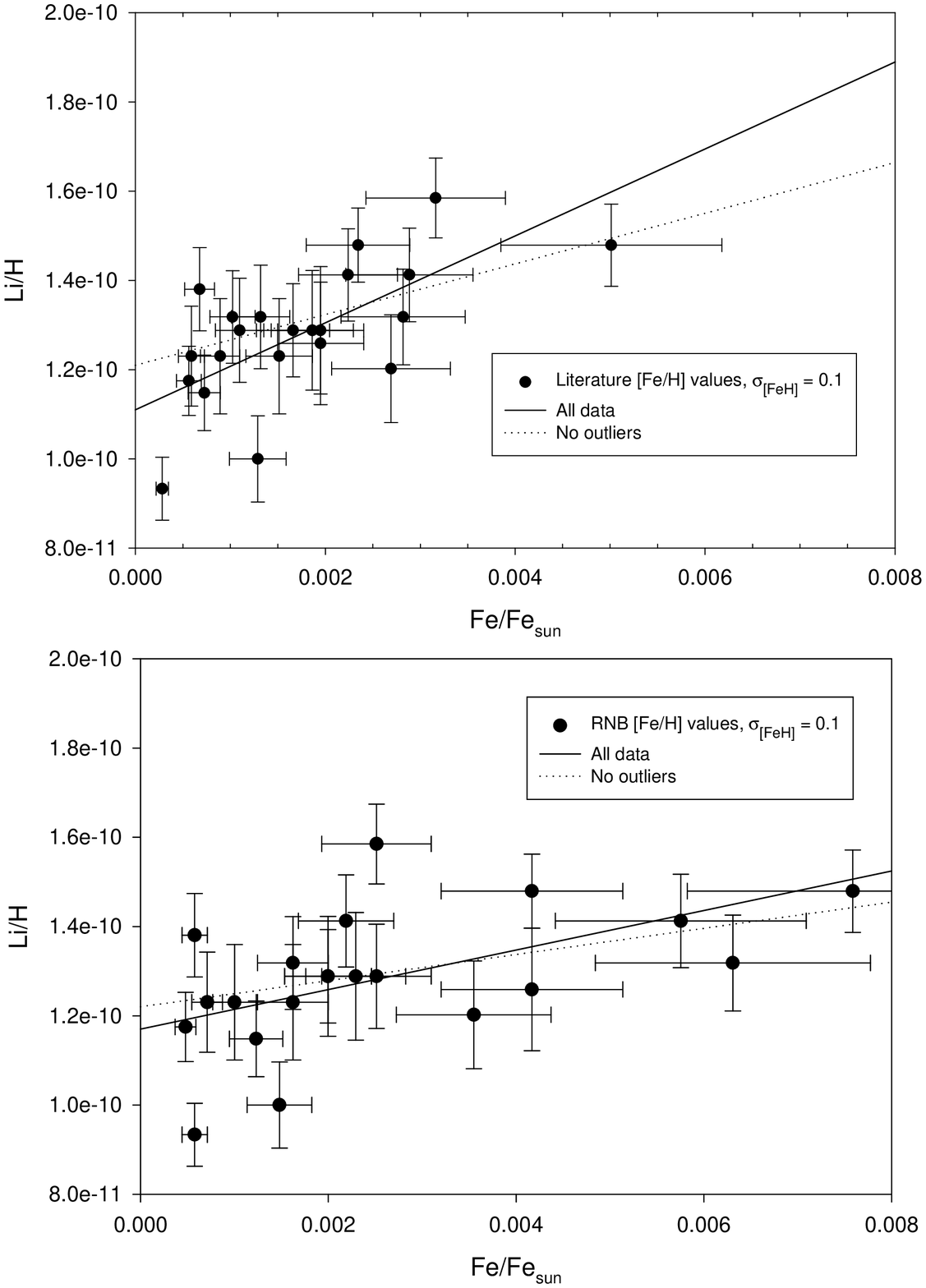}
\end{figure}

\begin{figure}
\plotone{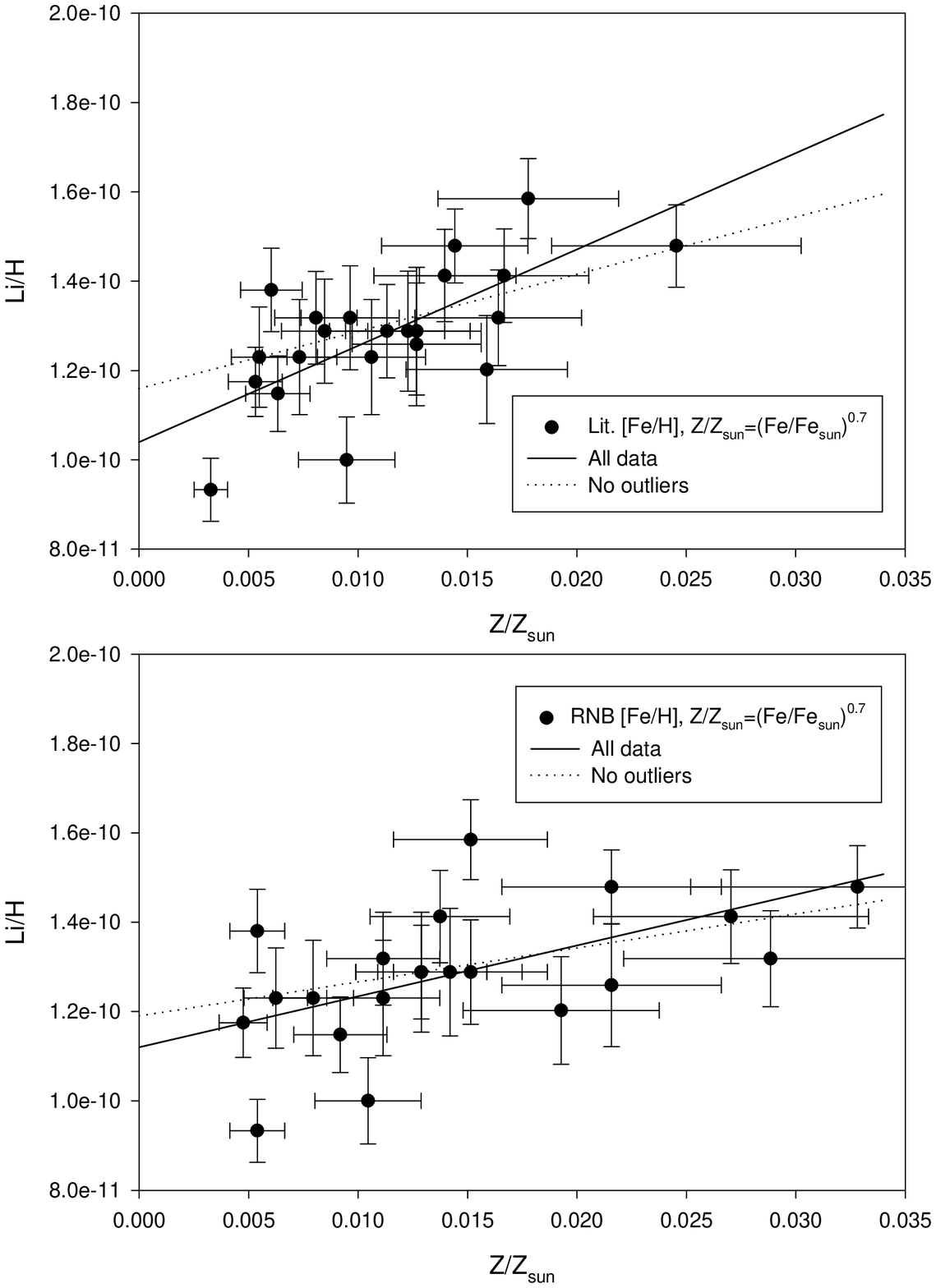}
\end{figure}

\begin{figure}
\plotone{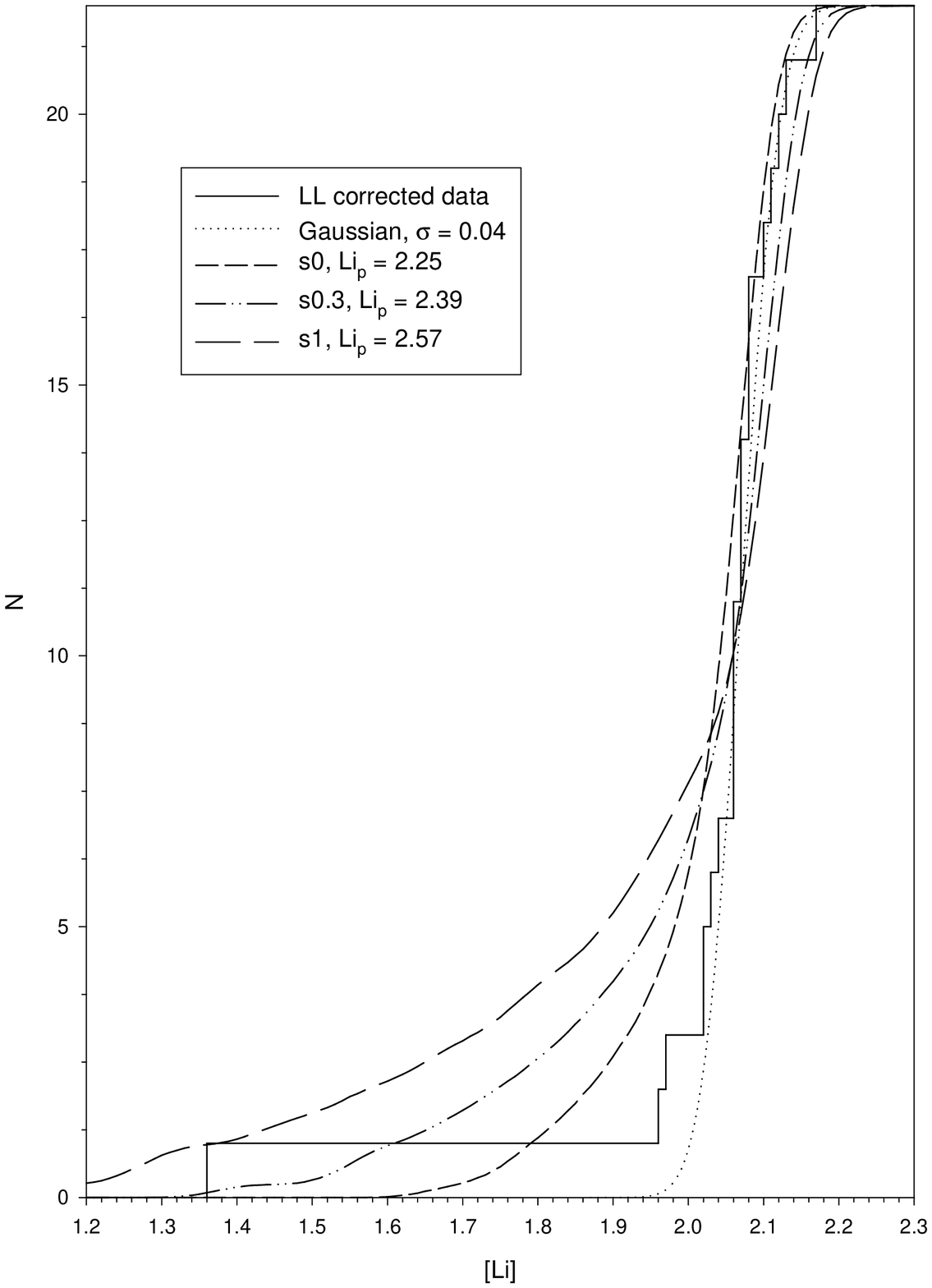}
\end{figure}

\begin{figure}
\plotone{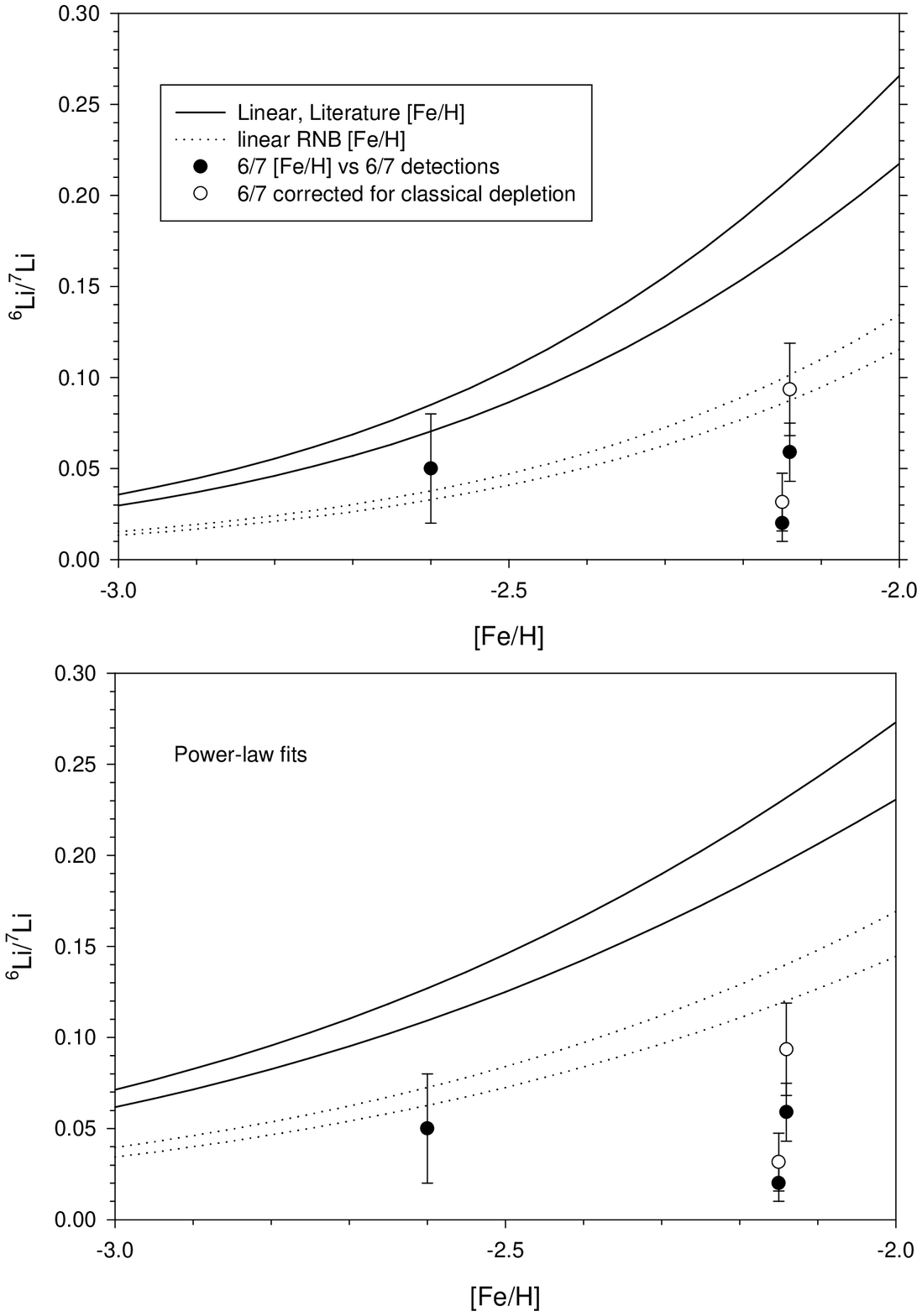}
\end{figure}

\begin{figure}
\plotone{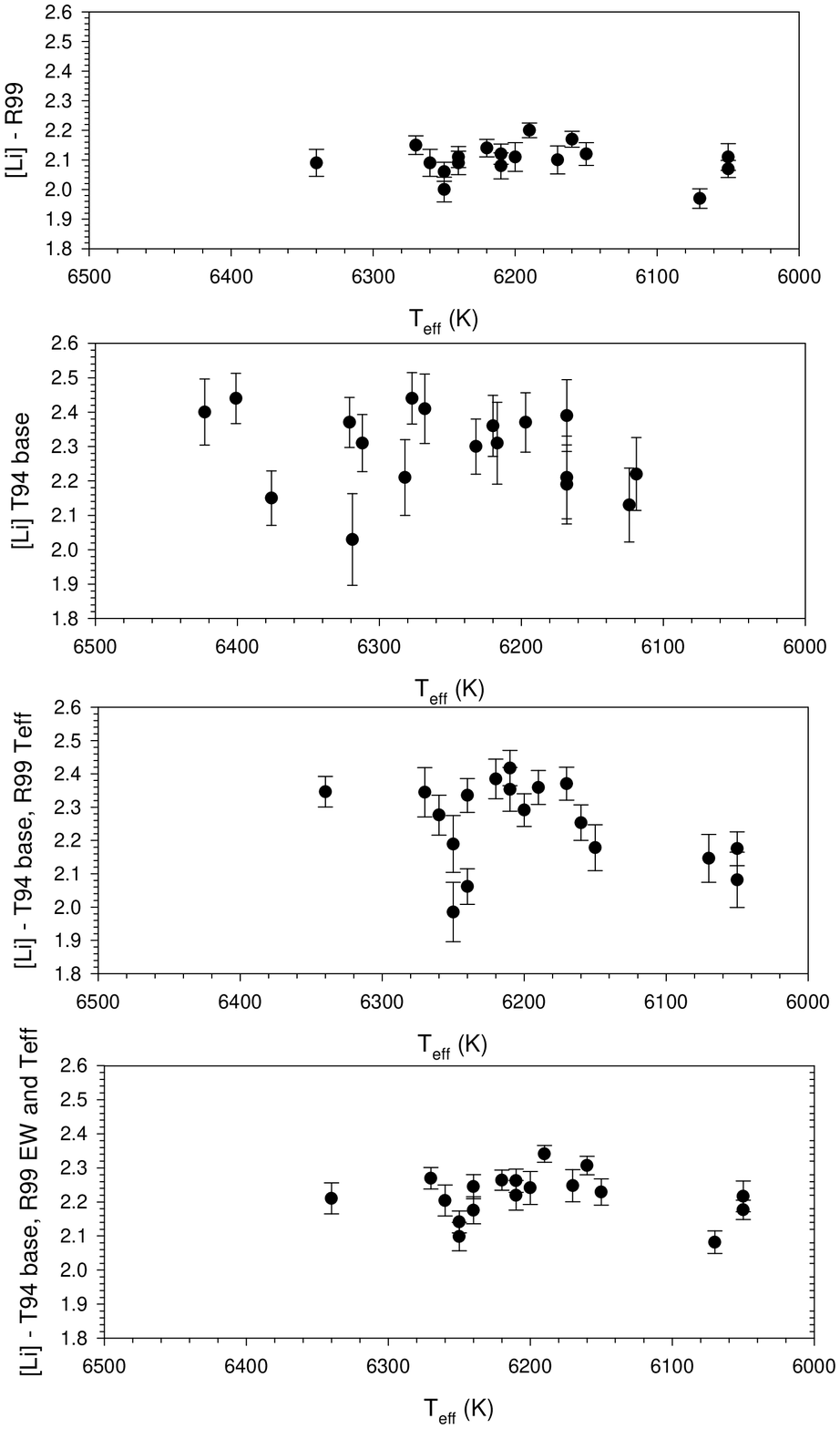}
\end{figure}

\begin{figure}
\plotone{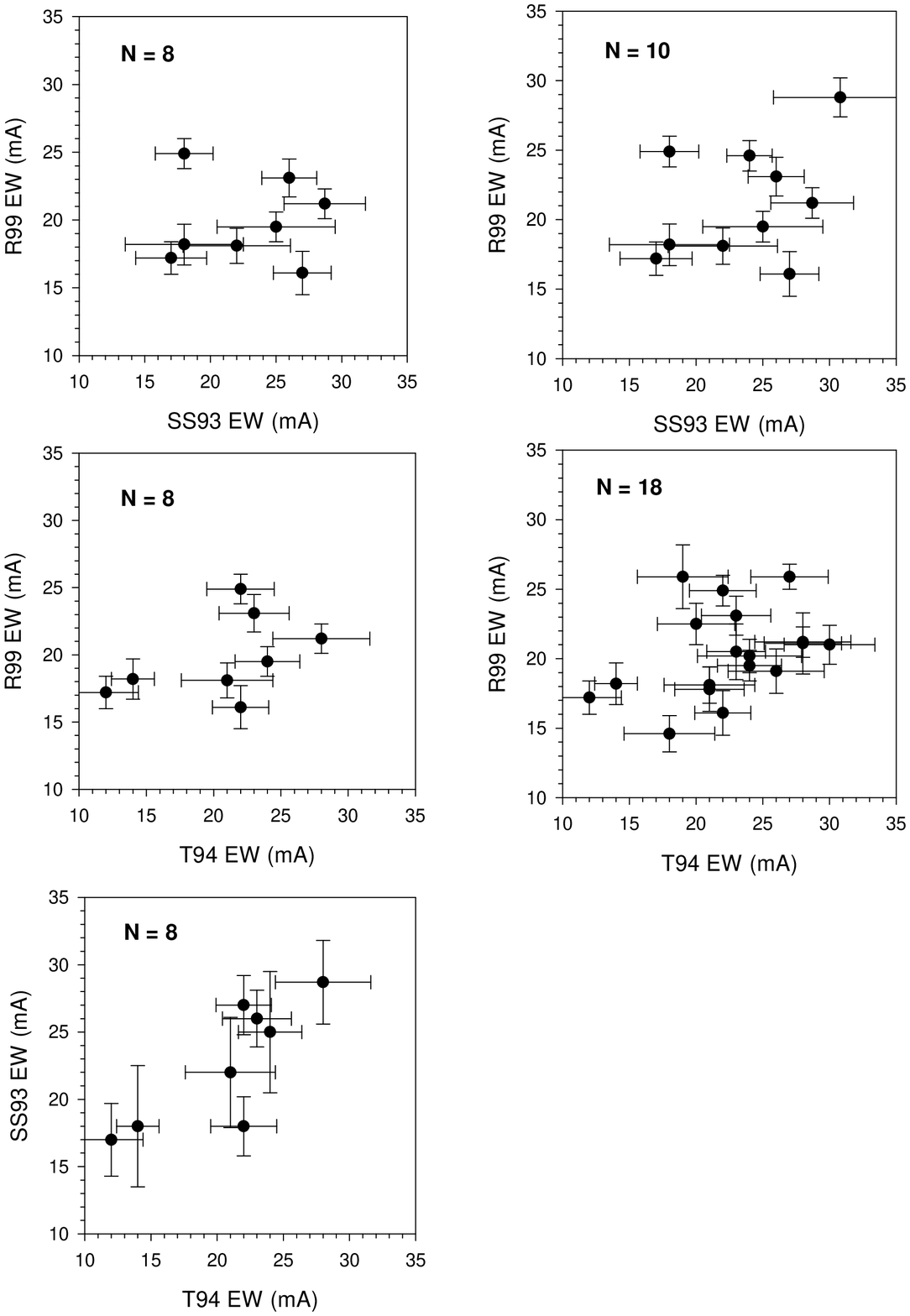}
\end{figure}

\begin{figure}
\plotone{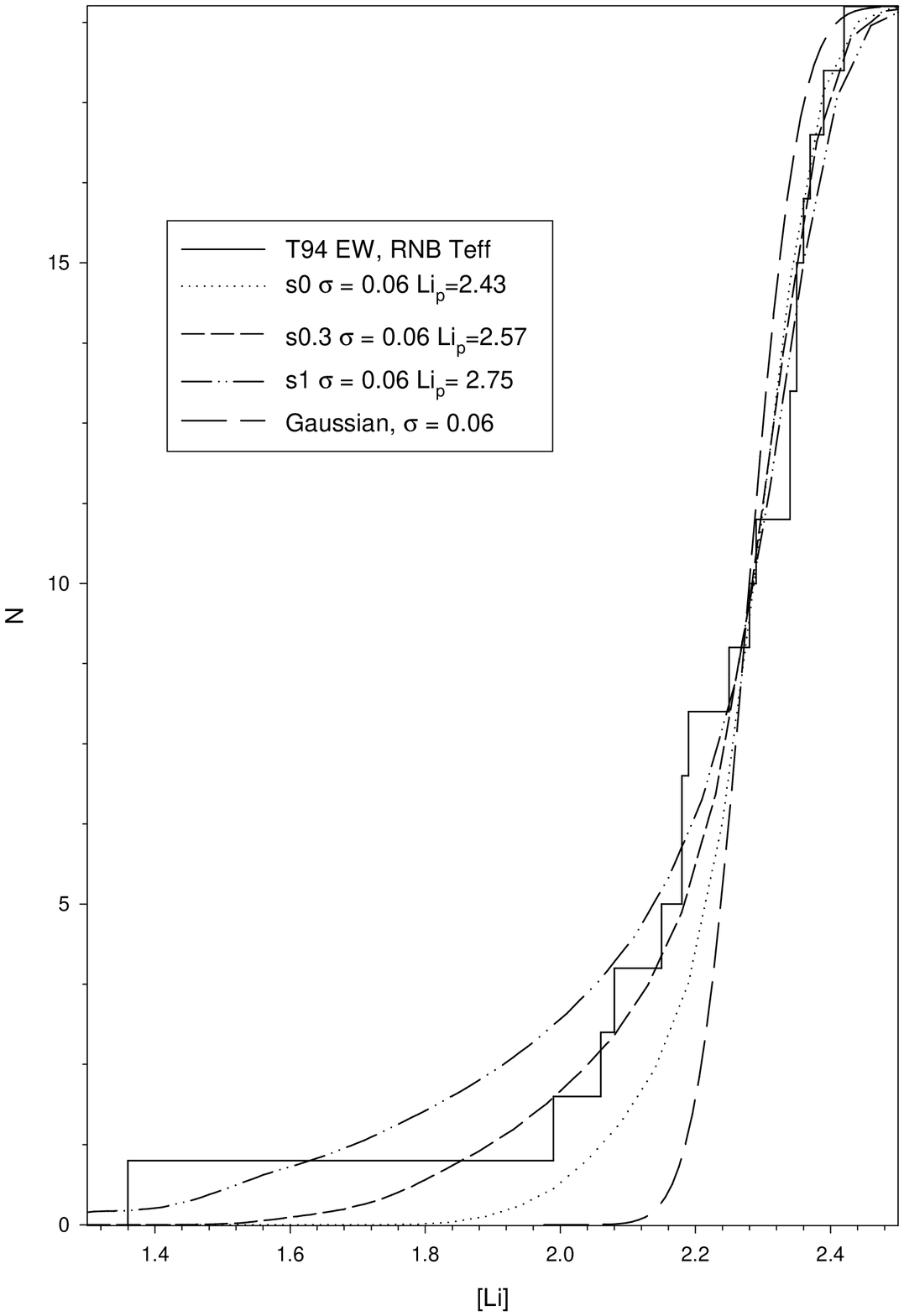}
\end{figure}

\begin{figure}
\plotone{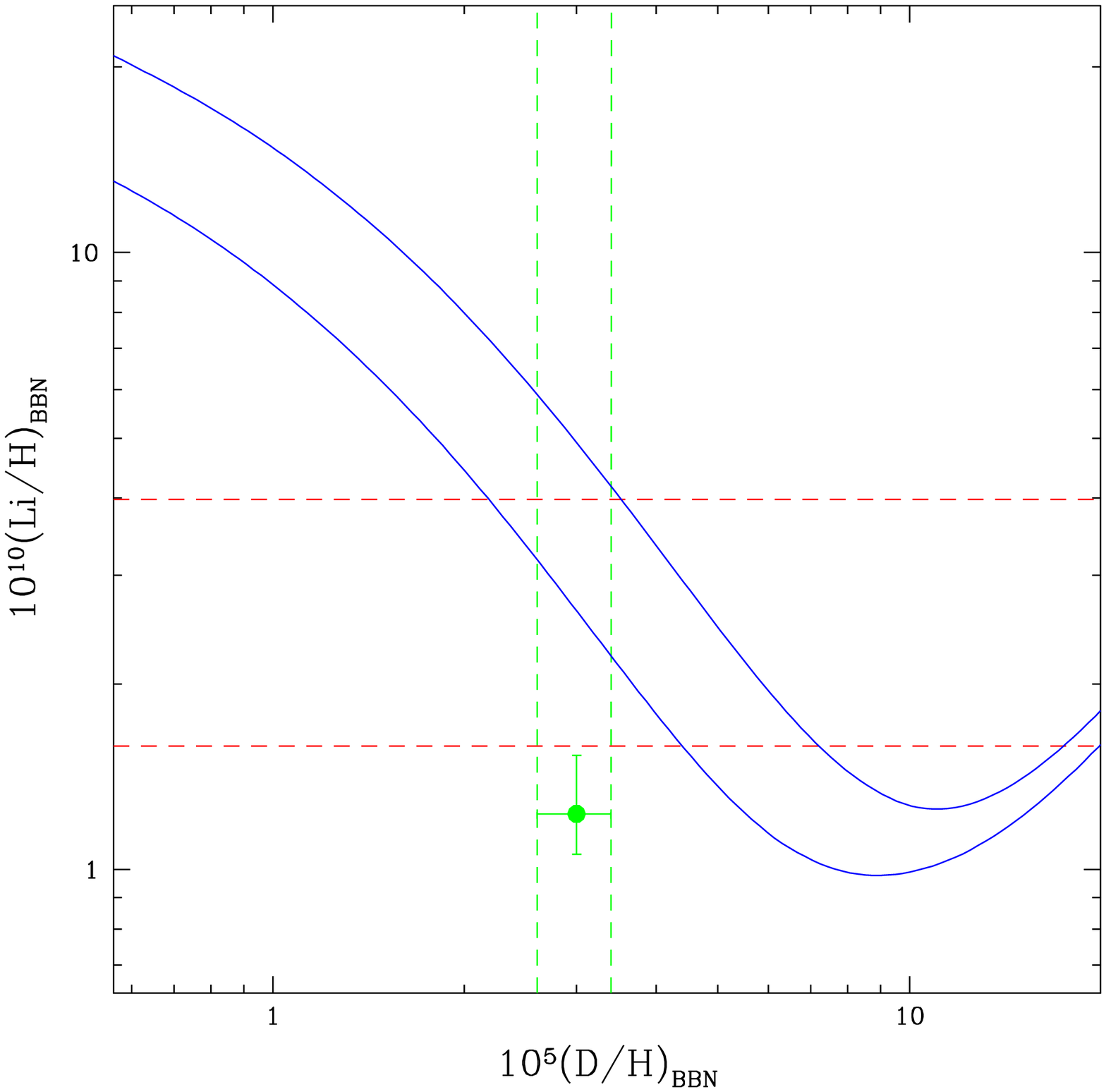}
\end{figure}

\end{document}